\documentclass[useAMS,usenatbib,a4]{mn2e}
\usepackage[]{graphicx}
\usepackage{epstopdf}

\title[The RSGs \& WR stars of NGC~604]{The red supergiants \& Wolf-Rayet stars of NGC~604}
\author[John J. Eldridge \& M\'onica Rela\~no]{John J. Eldridge$^{1}$ \thanks{E-mail: jje@ast.cam.ac.uk} \& M\'onica Rela\~no$^{1}$ \thanks{Email: mrelano@ast.cam.ac.uk} \\
$^{1}$Institute of Astronomy, The Observatories, University of Cambridge, Madingley Road, Cambridge, CB3 0HA\\}

\pagerange{\pageref{firstpage}--\pageref{lastpage}} \pubyear{2005}
\begin{document}
\maketitle
\label{firstpage}

\begin{abstract}
  We study the post-main sequence stars in NGC~604, the most luminous
  HII region in M33. Previously, a number of Wolf-Rayet (WR) stars and
  one red supergiant (RSG) have been discovered. Based on broadband
  photometry of the region, we present evidence that is consistent
  with the presence of this RSG and with three more RSG
  candidates. Using SED fitting based on HST UVIJHK photometry we
  estimate the ages of the WR stars and RSGs finding that the two
  populations are from distinct formation episodes with ages
  $3.2\pm1.0$\,Myrs and $12.4\pm2.1$\,Myrs, respectively. The RSGs
  have greater extinctions towards their line of sight than the WR
  stars consistent with the RSGs producing large amount of dust. Using
  the WR and RSG populations and similar SED fits to the most massive
  O stars we estimate that the total stellar mass is
  $(3.8\pm0.6)\times10^5\rm\,M_{\odot}$. We find a large discrepancy
  between the expected H$\alpha$ flux from such a massive cluster and
  that one observed. This suggests that $49^{+16}_{-19}$ percent of
  the ionizing photons produced by massive stars in NGC~604 is leaking
  from this HII region. We also suggest that the implications of an
  old RSG population mean that if NGC~604 was more distant and only
  observed in the infrared (IR) it would be difficult to study the
  youngest burst of star formation due to the contamination of RSGs.
\end{abstract}


\section{Introduction}
NGC~604 is the most luminous and massive HII region in the nearby
galaxy M33. Due to its large size, great mass and proximity it has
been intensively studied to gain insight into the physics of HII
regions and the evolution of massive stars \citep[e.g.][among
  others]{hunter,terlevich,drissen93,ngc604wr,maiz,rosa}.  Of
paramount importance is to determine the age of the stellar population
and the different stellar types present in this region, because a
deeper knowledge of this object can help us to infer properties for
star-forming regions of similar types in distant galaxies.

The confirmed stellar constituents of NGC~604 are OB stars, WR stars
and at least one spectroscopically confirmed RSG
\citep{hunter,terlevich,ngc604wr}. The different stellar types that
NGC~604 harbours makes this HII region an ideal object to study its
stellar population and to test stellar evolution and population
synthesis models. NGC~604 is a key region to test whether the
population inferred from \textit{resolved} stellar populations can
also match the \textit{unresolved} spectral features that would be
used in more distant galaxies. In this paper we use this region to
qualitatively check the accuracy of the binary population and spectral
synthesis (BPASS) code described in \citet{es09} and at
\texttt{http://www.bpass.org.uk}. Here we analyze the resolved stellar
population observed with deep Hubble Space Telescope (HST) images and
study the region as an unresolved object by using the integrated
spectrum \citep{rosa} and observations of \citet{relano}.

The possible presence of RSGs in NGC~604 comes from the spectroscopic
evidence of \citet{terlevich}. \citet{hunter} suggested the existence
of a supergiant population based on HST stellar photometry, but these
authors could not elaborate further these ideas since their
observations where limited to the visible part of the
spectrum. Recently, \citet{barba} have shown the complexity of the
stellar population of this region using infrared NICMOS (Near Infrared
Camera and Multi-Object Spectrometer) observations. They found
evidence of massive young stellar object candidates and suggested the
existence of a RSG population within the region.

Here we aim to study in detail the post-main sequence objects RSGs and
WR stars of NGC~604, along with the most massive main-sequence stars
in the region. We obtain the photometry for the stars within NGC~604
in the IR-NICMOS filters and combine the results with the optical
photometry performed by \citet{hunter}. We analyze the Spectral Energy
Distribution (SED) of these stars from the optical to the IR
wavelength range and investigate whether the RSG and WR populations
have different ages. We then attempt to infer from these observations
the total stellar mass and check if this is consistent with other
observations such as the total H$\alpha$ flux from NGC~604.

\section{Evidence of Red Supergiants in NGC~604}

The stellar content of NGC 604 was previously studied by Hunter et al.
(1996) using UVI photometry. These authors showed the existence of
numerous massive main-sequence stars and identified a candidate WR
population in their colour-magnitude diagram (CMD). They also suggest
the existence of numerous stars bright enough to be blue or red
supergiants, which would indicate an older sub-population in
NGC~604. Due to the limitation of their data, restricted only to the
visible part of the spectrum, they did not study this cool and evolved
stellar population. Taking advantage of the available NICMOS-IR data
for this region and the comparison with the optical photometry, we now
investigate further this evolved population. Using Hunter et al.'s
photometry we selected stars in the central cluster of the region
(Cluster A, as defined in \citet{hunter}) with magnitudes and colours
in the range of F555W-F814W$>$1 and F555W$<$-6 and make a list of
possible RSG candidates in NGC~604. These limits ensure we are taking
massive ($\rm M>15\rm\,M_{\odot}$) RSGs and therefore we can compare
this population with the massive WRs detected by \citet{ngc604wr}.

\subsection{NICMOS data analysis}

NGC~604 was observed with the NICMOS (NIC2) on board the Hubble Space
Telescope (HST) under the ID program 10419 (IP: R. Barb\'a) with the
filters F110W, F160W and F205W (similar to J, H and K filters). We
retrieved 6 fields from the HST Multimission Archive at the Space
Telescope Institute (MAST), 5 covering the central part of NGC~604 and
another covering an outside field for background subtraction. Each
image has a field of view of 19.2 $\times$19.2 arcsec and a pixel
scale of 0.075 arcsec.  

\begin{figure*}
\includegraphics[angle=0, width=164mm]{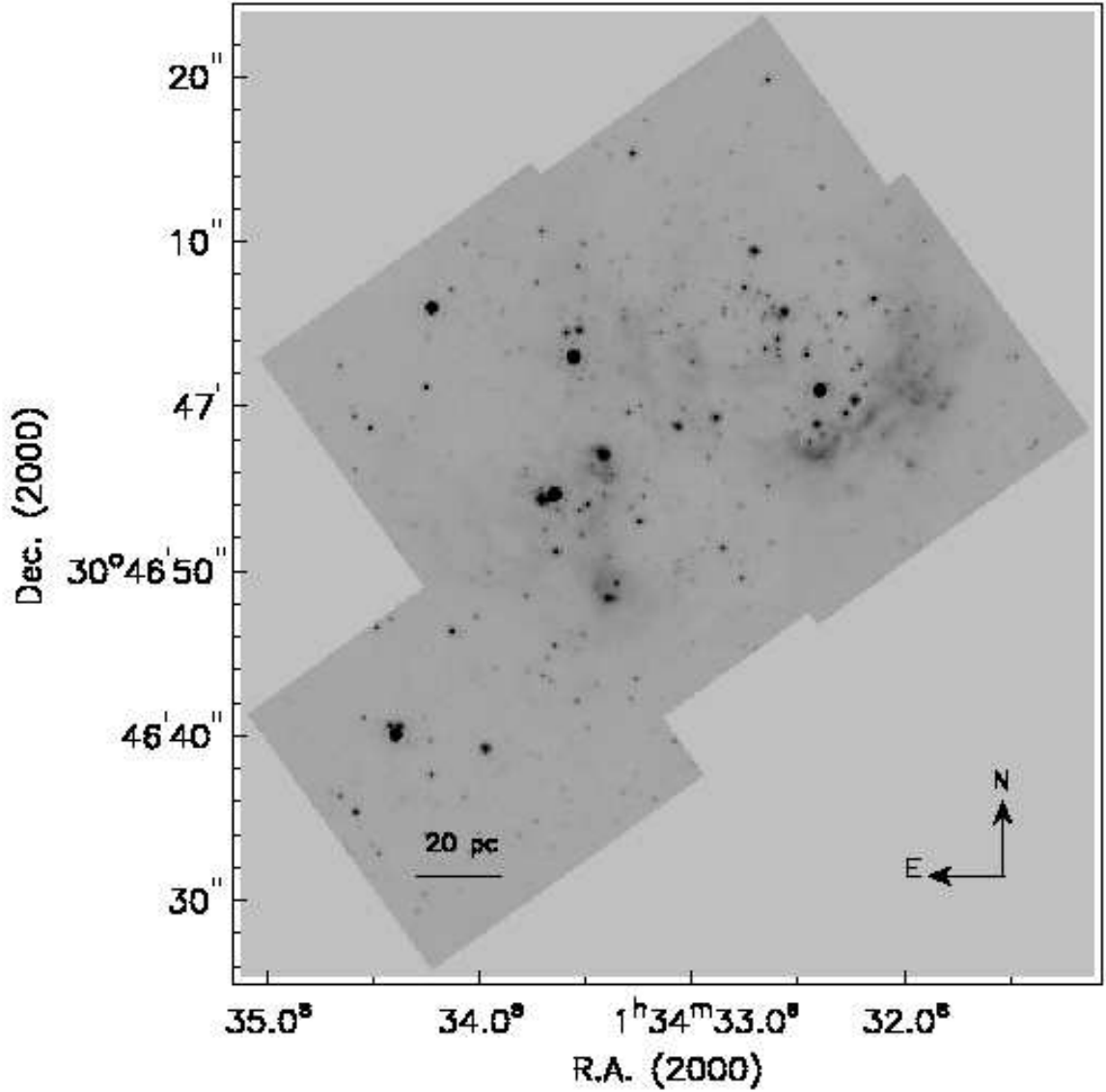}
\caption{Final mosaic F205W image of NGC~604 obtained with
  NICMOS-NIC2 Camera. The total field of view of NICMOS observations is 50''$\times$60''
  with a pixel scale of 0.075 arcsec.}
\label{F205Wa}
\end{figure*}

The retrieved data were processed by the pipeline software Calnica
(Version 4.4.0) but further analysis needed to be performed to obtain
the final mosaic images. First we, masked out the Coronagraphic Hole
and applied the MultiDrizzle software package\footnote{$\rm
  http://stsdas.stsci.edu/multidrizzle$} to create the final cosmic
ray-free mosaic images.  The retrieved F205W images showed some
residuals due to a bad flat-field correction after being processed by
Calnica with the default calibration files provided in the data
archive. Therefore we generated a new flat-field using the F205W
background images and reprocessed the images with it. The use of this
flat-field clearly improved the final images for this filter. Finally,
the $\sim$20 rows of the CCD affected by vignetting were eliminated
from the images. We then created the mosaic image for the 5 fields in
each filter using the World Coordinate System information from the
headers. The final image for the F205W filter is shown in
Fig.\,\ref{F205Wa}. The angular resolution (FWHM) in the final images
are 0.16, 0.18, 0.20\,arcsec for the corresponding filters F110W,
F160W, F205W.

We obtained the stellar magnitudes for each filter using the
crowded-field fitting software DAOPHOT \citep{stetson}.  We identified
stars with intensities above 5 times the r.m.s. noise of each image
using DAOFIND, then aperture photometry was performed on the selected
stars with a circular aperture of 3 pixel radius (0.23\,arcsec). A
Point-spread function (PSF) was fitted using bright and isolated stars
in each image. Then, the instrumental magnitude was derived for each
identified star using the fitted PSF in each image. We subtracted the
identified stars off the original image and we run DAOFIND again in
the residual image to select fainter stars that were not identified in
the original images. The magnitudes of these new stars were derived
and included in the final photometry. In this second iteration of
DAOPHOT we detected the following stars in each filter, 22 stars in
F205W, 41 stars in F160W and 47 stars in F110W. However, we are only
modelling stars detected in the 3 NICMOS filters. With this condition
the second iteration add only 5 stars to the sample in the
models. None of these 5 stars are in the group classified as WR, RSG
or massive OB stars, and therefore it does not have impact in the
analysis nor conclusions presented in the paper. Aperture corrections
to a radius of 0.49\,arcsec were applied to the instrumental
magnitudes in order to use the photometric calibrations recorded in
the image headers and provided by the NICMOS team\footnote{$\rm
  http://www.stsci.edu/hst/nicmos/performance/photometry$}.  Finally,
magnitudes in the Vega system for each star and each filter were then
obtained.

\begin{figure*}
\includegraphics[width=\textwidth]{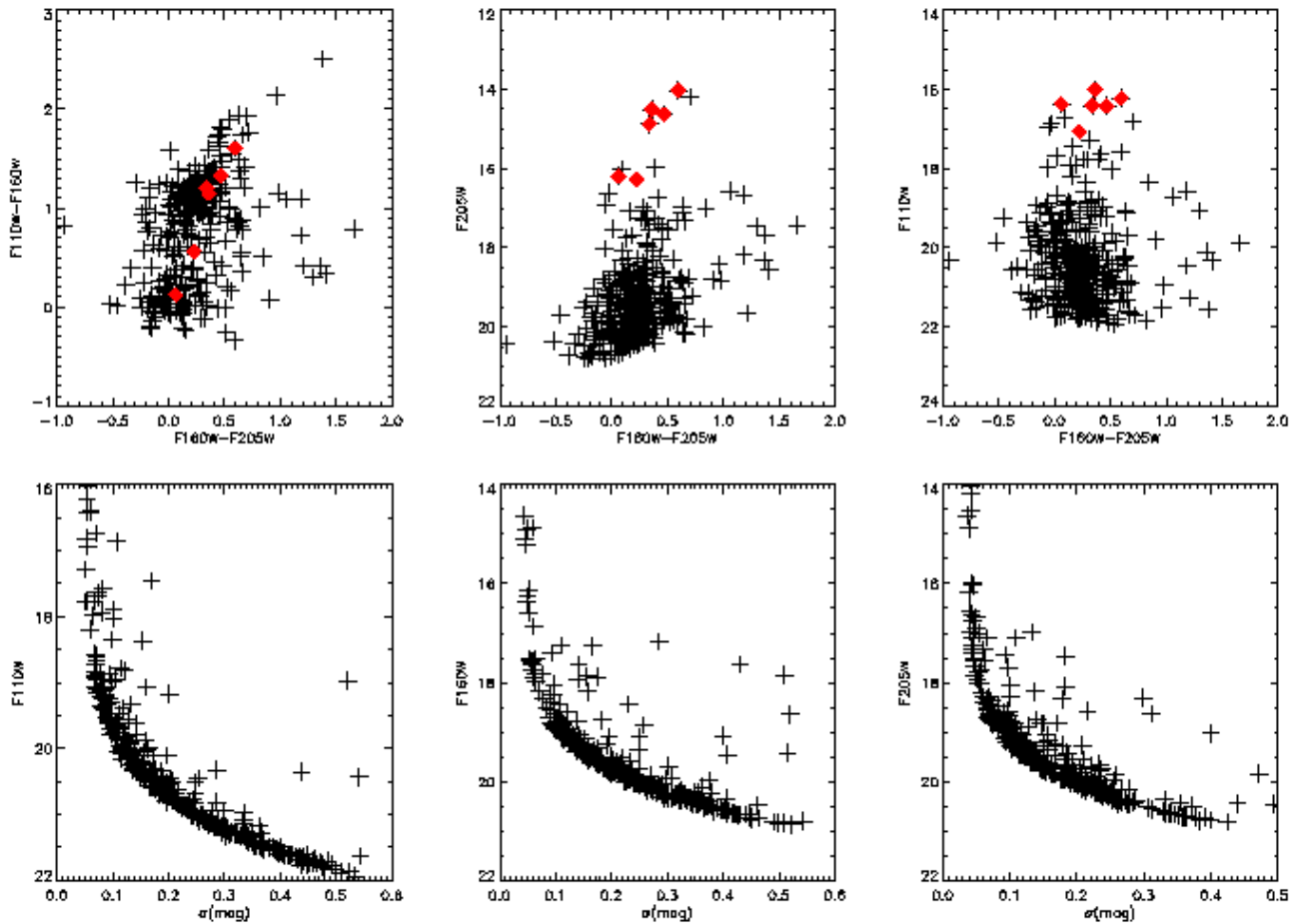}
\caption{Top panel: Colour-colour diagram (left) and CMDs (centre and
  right) for NGC~604 in the NICMOS filters F110W, F160W and F205W. The
  red diamonds are the candidate RSG stars selected using the UVI
  stellar photometry from \citet{hunter}. Bottom panel: Observed
  magnitude as a function of the estimated errors for each filter:
  F110W (left), F160W (centre) and F205W (right). No extinction
  corrections have been applied to the magnitudes and all the
  magnitudes are in the Vega system.}
\label{mag}
\end{figure*}

The result of the stellar photometry is shown in the top panels of
Fig.\,\ref{mag}. We present here the colour-colour and CMD for the
identified stars in NGC~604. The magnitude errors are a combination of
the image photon noise and the uncertainties derived from the PSF
fitting procedure and from aperture corrections; these are shown as a
function of the magnitude for each filter in the lower panels of
Fig.\,\ref{mag}. The red diamonds in the top panels of Fig.\,\ref{mag}
correspond to the RSG candidates derived from the UVI photometry from
\citet{hunter}.  We identified the RSG candidates in the NICMOS images
using the coordinates of Hunter et al.'  star catalogue and the F555W
image from their study. Only 6 candidates were within the field of
view of NICMOS. With exception of one RSG candidate (number 2 in
Table\,\ref{obsstars}), for which we identified another star very
close to it with magnitudes in all NICMOS filters and with no counter
part in the UVI photometry, the rest of the stars are clearly
identified as single stars in the NICMOS images. For the star 2 in
Table\,\ref{obsstars} we added the IR fluxes of the two identified
stars in order to perform the SED fitting with all the filters
available (see section~3).

\subsection{Confusion with foreground stars}
The identification of RSGs in extragalactic clusters is a difficult
task as they could be confused with foreground Galactic stars. As it
has been explained in \citet{massey98}, a 40~$\rm\,M_{\odot}$ RSG in
M33 would have an apparent magnitude in V-Band between $+$15.5 to
$+$18, depending on its effective temperature. For dwarf late-type
stars the absolute V magnitude is $+$6 (for a K0 dwarf star) and $+$12
(for a M5), taking $\sim$~600~pc as the estimated distance for these
stars in the direction of M33, the apparent V magnitudes for these
stars will be between $+$15 and $+$21 \citep{massey98}. Therefore, it
is possible to mistake a Galactic dwarf late-type star for a RSG in
M33.
 
\begin{figure}
\includegraphics[width=0.5\textwidth]{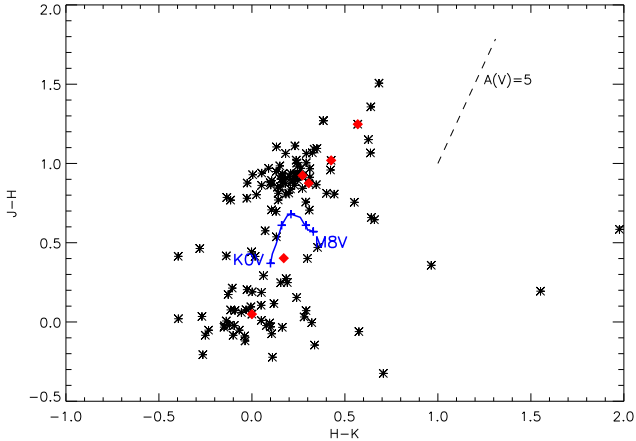}
\caption{Colour-colour diagram in the JHK system for the stars
  identified in NICMOS images and magnitude errors less than
  0.2\,mag. Red diamonds correspond to the RSG candidates shown in
  Fig.\,\ref{mag}.  The blue line shows the location of foreground
  late-type dwarf stars. The transformation from the NICMOS filter to
  JHK system has been performed following the equations given in
  \citet{brandner01}. The reddening vector corresponds to Av=5mag and
  the \citet{cardelli} extinction law. }
\label{fore}
\end{figure}

A method of separating foreground stars from RSGs based on IR
two-colour diagrams have been proposed by \citet{elias85}: the RSG are
0.1-0.2 mag redder than late-type dwarfs in J-H at comparable H-K. In
Fig.\,\ref{fore} we show the JHK colour-colour diagram for the stars
with NICMOS photometry as well as for the RSG candidates. The
magnitudes in the NICMOS filters have been converted into magnitudes
in the JHK system following the transformation of \citet{brandner01}.
Since the magnitude conversion procedure already adds uncertainties in
the magnitude we only convert those stars with magnitudes errors lower
than 0.2\,mag. The blue line denotes the location of the foreground
late-type dwarf stars with magnitudes obtained from
\citet{Koornneef83}. Clearly four of our RSG candidates stand between
0.2 to 0.6 magnitudes above the location defined by the foreground
stars. The two stars that lie below the line defined by the foreground
stars are S2 and S5. These stars, as we will see later, have anomalous
SED fitting (see section~4).

\section{Stellar evolution models and SED fitting}

Determining the physical parameters of stars from photometry requires
a combination of stellar evolution models, model stellar atmospheres
and a method to find the best fit between models and theory. In this
section we outline the stellar and atmosphere models, their
amalgamation and how we use the results to estimate parameters for the
observed stars. To calculate the absolute magnitudes of the observed
stars we take the distance of NGC~604 to be 840 kpc \citep{Freedman}.

We use stellar models from the Cambridge STARS code \citep[][and
  references therein]{EIT} that have been correlated with stellar
atmosphere models to produce synthetic spectra and broad-band
photometric colours as described in \citet{es09}. Their key feature is
that there is not only a set of detailed single star models, but also
an extensive set of detailed binary star models which are key to
producing a realistic synthetic stellar population. We consider
stellar models at the relevant metallicity for NGC~604 Z=0.008
\citep{vilchez88,esteban09}, where Z is the metallicity mass fraction
(a metallicity of Z=0.020 is conventionally considered solar). It is
well known that local metallicities within disk galaxies can vary by
significant factors away from the mean host metallicity. We inspected
the affect of varying metallicity by running the fits with models at
metallicities of $Z=0.004$ and $0.020$. We find that our results
change only slightly and allowing the metallicity to vary would only
increase the uncertainty of our results. We assume a hydrogen mass
fraction of $X=0.73$ and a helium mass fraction of $Y=0.262$. We
calculate observable magnitudes for these stellar interior models by
using a model atmosphere that best represents the appearance of the
interior model at each timestep. The atmosphere models are for OB
stars, WR stars and other stellar types and they are described in
\citet{crow}, \citet{wn} and \citet{basel}, respectively.

Given that stellar evolution is non-linear, and binary evolution is
even more so, we do not interpolate between models with different
initial parameters to determine the nature of the observed
stars. Instead, we compare the observed SEDs to the synthetic SEDs at
every timestep of each stellar model and estimate a most likely mass,
age and extinction for each observed star. This is done by calculating
mean parameters from all the models that match the observed SED.

We first estimate the amount of extinction required for the synthetic
SED to match the observed SED. We calculate the amount of extinction
required in each band to achieve an exact match between the observed
and model magnitudes. Using the \citet{cardelli} extinction curve we
convert each magnitude to the equivalent extinction in $A_v$
calculating the conversion factor $\alpha_i$, such that $A_i=\alpha_i
A_v$ for the $i$th filter. We then calculate the mean extinction from
the six filters, as follows,
\begin{equation}
A_{v,n}(m,t) = \frac{1}{N} \sum_{i=1}^{N} \frac{(M_{{\rm obs},n,i}-M_{{\rm model},i}(m,t))}{\alpha_i},
\end{equation}
where the magnitudes predicted by a model are $M_{{\rm model},i}(m,t)$
for a star with initial mass $m$ and age $t$, the observed magnitude
is $\rm\,M_{{\rm obs},n,i}$ for the $n$th star in the $i$th filter and
$N$ is the number of filters in which the source is detected. This
mean extinction is then used to correct the synthetic SED magnitudes
for dust.

This dust corrected model magnitudes, 
\begin{equation}
M_{{\rm model,cor},i,n}(m,t)=M_{{\rm model},i}(m,t)+\alpha_i A_{v,n}(m,t),
\end{equation}
are then used to calculate the probability, $p_n(m,t)$,
of a match between the observed SED and theoretical SED can be
evaluated as follows:
\begin{equation}
p_n(m,t)= \prod_{i=1}^{6} \exp\Big(-\frac{(M_{{\rm model,cor},i,n}(m,t) - M_{{\rm obs},n,i})^2}{2\sigma_{{\rm total},i}^2}\Big)
\end{equation}
The errors are therefore assumed to be Gaussian. $\sigma_{\rm total}$
is calculated as a combination in quadrature of the error in the
photometry and an assumed error in the stellar models of 0.3 mag. This
error is estimated from the magnitude difference between two stars of
neighbouring masses in our grid of models. We must also take into
consideration the resolution of our mass grid in the probability
calculation. For example our $15M_{\odot}$ model has neighbouring model
masses of $14$ and $16M_{\odot}$. While the $80M_{\odot}$ model has
neighbours in the mass grid of $70$ and $100M_{\odot}$. Therefore we
add a factor, $\Delta m(m)$, to the probability to give every mass an
equal probability. For the the 15 and $80M_{\odot}$ examples this
factor would be 1 and 15 respectively.

The last factor to consider in the likelihood of a match is how
quickly stars evolve. Two stars might have the same apparent SED but
one is evolving more rapidly than another. A star evolving slowly,
such as main-sequence stars will be more probable matches to a SED
that a short-lived WR stars. Therefore we use the probability and the
timestep of each model SED as a weight to estimate values for the
parameters of each observed star. For example, the mass ($M_{\rm n}$),
age ($\tau_{\rm n}$) and extinction ($A_{\rm v, n}$) are obtained as
follows:

\begin{equation}
M_{\rm n}= \frac{\sum_{m}^{\rm all}\sum_{t}^{\rm all} m \, \Delta t(m,t) \, \Delta m(m) \,  p_n(m,t)}{\sum_{m}^{\rm all}\sum_{t}^{\rm all} \Delta t(m,t) \, \Delta m(m) \,  p_n(m,t)}\\
\end{equation}

\begin{equation}
\tau_{\rm n}= \frac{\sum_{N}^{\rm all}\sum_{t}^{\rm all} t \,  \Delta t(m,t) \, \Delta m(m) \,  p_n(m,t)}{\sum_{m}^{\rm all}\sum_{t}^{\rm all} \Delta t(m,t) \, \Delta m(m) \,  p_n(m,t)}\\
\end{equation}

\begin{equation}
A_{\rm v, n}= \frac{\sum_{m}^{\rm all}\sum_{t}^{\rm all} A_{\rm v,n}(m,t) \,  \Delta t(m,t) \, \Delta m(m) \,  p_n(m,t)}{\sum_{m}^{\rm all}\sum_{t}^{\rm all} \Delta t(m,t) \, \Delta m(m) \,  p_n(m,t)}\\
\end{equation}
and for each of these we are able to estimate an error from the variance, for example:
\begin{equation}
\sigma_{M,n}^2= \frac{\sum_{m}^{\rm all}\sum_{t}^{\rm all} m^2 \,  \Delta t(m,t) \, \Delta m(m) \,  p_n(m,t)}{\sum_{m}^{\rm all}\sum_{t}^{\rm all} \Delta t(m,t) \, \Delta m(m) \,  p_n(m,t)}-M_{\rm n}^2\\
\end{equation}
Thus for each observed SED we obtain an estimate of age, initial mass,
extinction, luminosity, surface temperature, the mass ratio
($q=M_2/M_1$) for the binary models and an error for each value. We
note that we are only determining three independent parameters from
each fit: luminosity, surface temperature and extinction. The age and
mass are model dependent and linked to the surface temperature and
luminosity. If we re-ran the analysis with different or updated models
the luminosity, surface temperature and extinction would remain
similar but the age and mass would change by a greater degree.

We perform our fitting process using two sets of models: single star
models and binary models, then we compare the probability from the two
fits to determine which is more likely. We find, except for the RSG1,
4 and 6, the binary models give a \textit{better} match for all the
stars we fit. This is likely to be due to the fact the binary models
can cover more of the Hertzsprung-Russel diagram than the single stars
therefore have a greater chance of achieving a close match to the
SED. In most cases the likelihood difference between single and binary
models is minimal up to only factors between 1 to 3. The only two
fits for which binaries are significantly favoured are RSG3 and
WR1. For these two stars they are 700 and 30 times more likely to be
binaries respectively.

The SED fits for the RSGs and WR stars assume that the stars are
post-main sequence objects and therefore we only use post-main
sequence models to prevent the fits being affected by main-sequence
models skewing the fit. This is because in weighing by the timestep a
WR star and a main-sequence OB star might have very similar SEDs but a
star is more likely to be a main-sequence star due to its slow
evolutionary timescale.

\begin{figure*}
\includegraphics[angle=0, width=58mm]{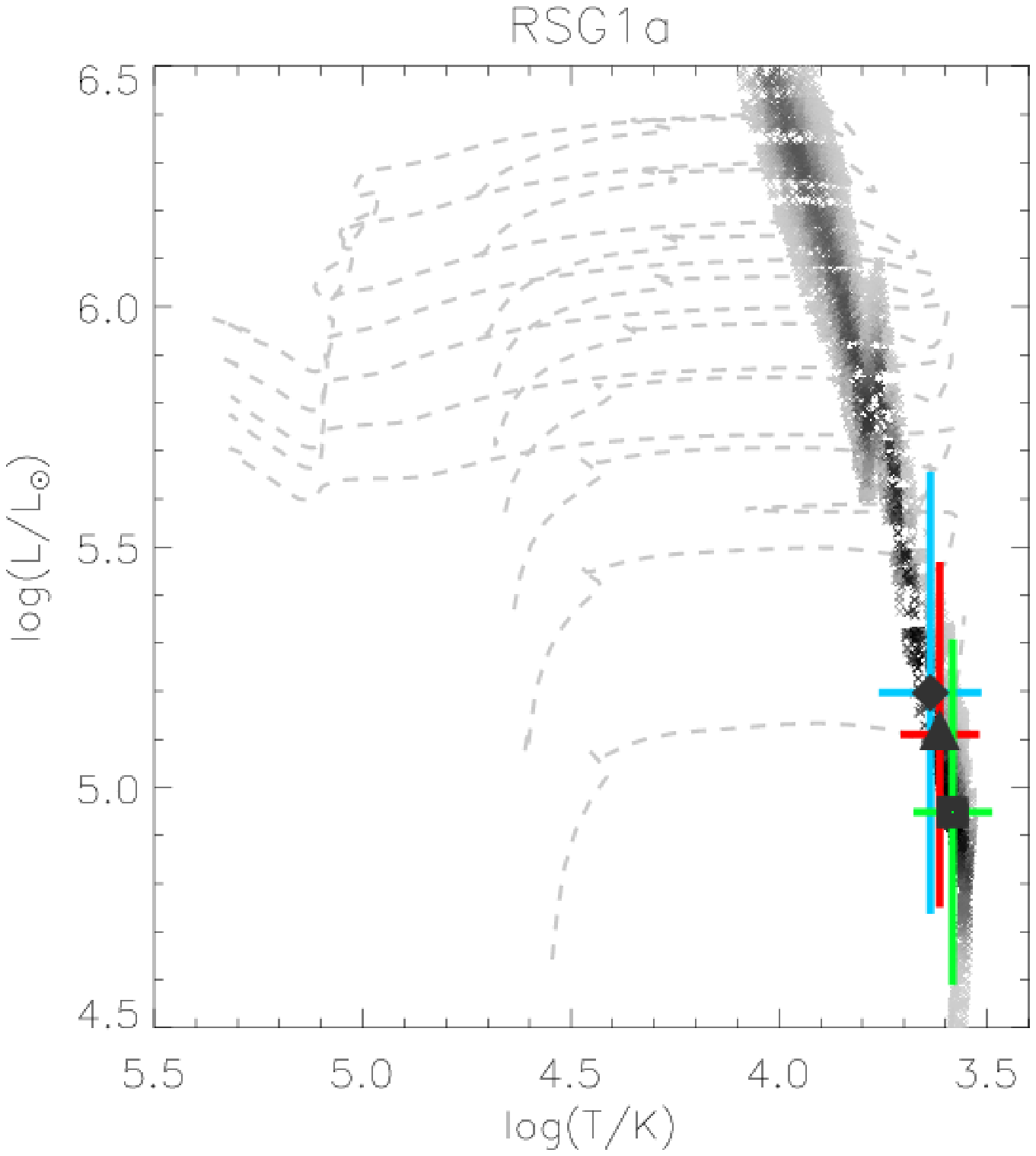}
\includegraphics[angle=0, width=58mm]{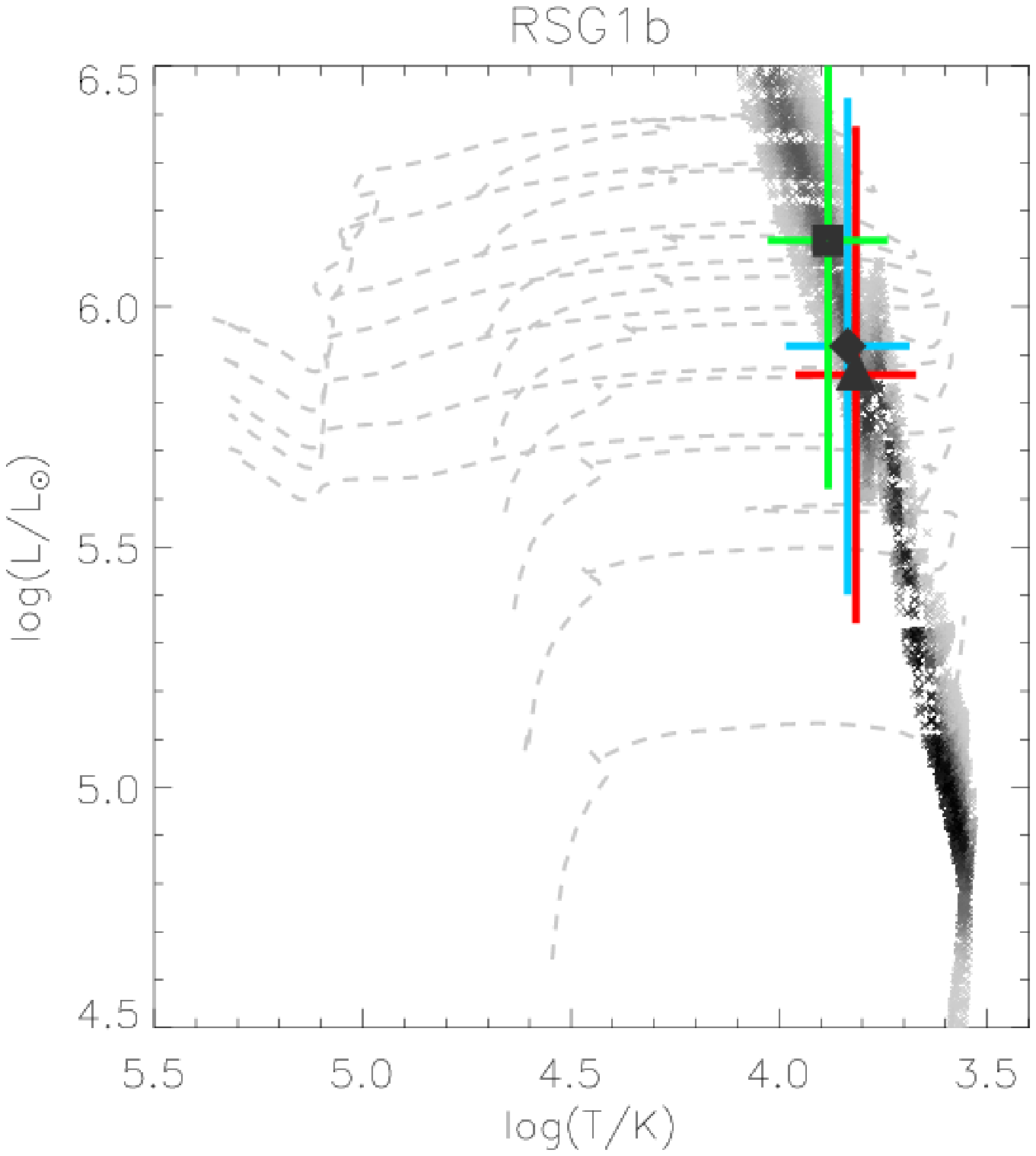}
\includegraphics[angle=0, width=58mm]{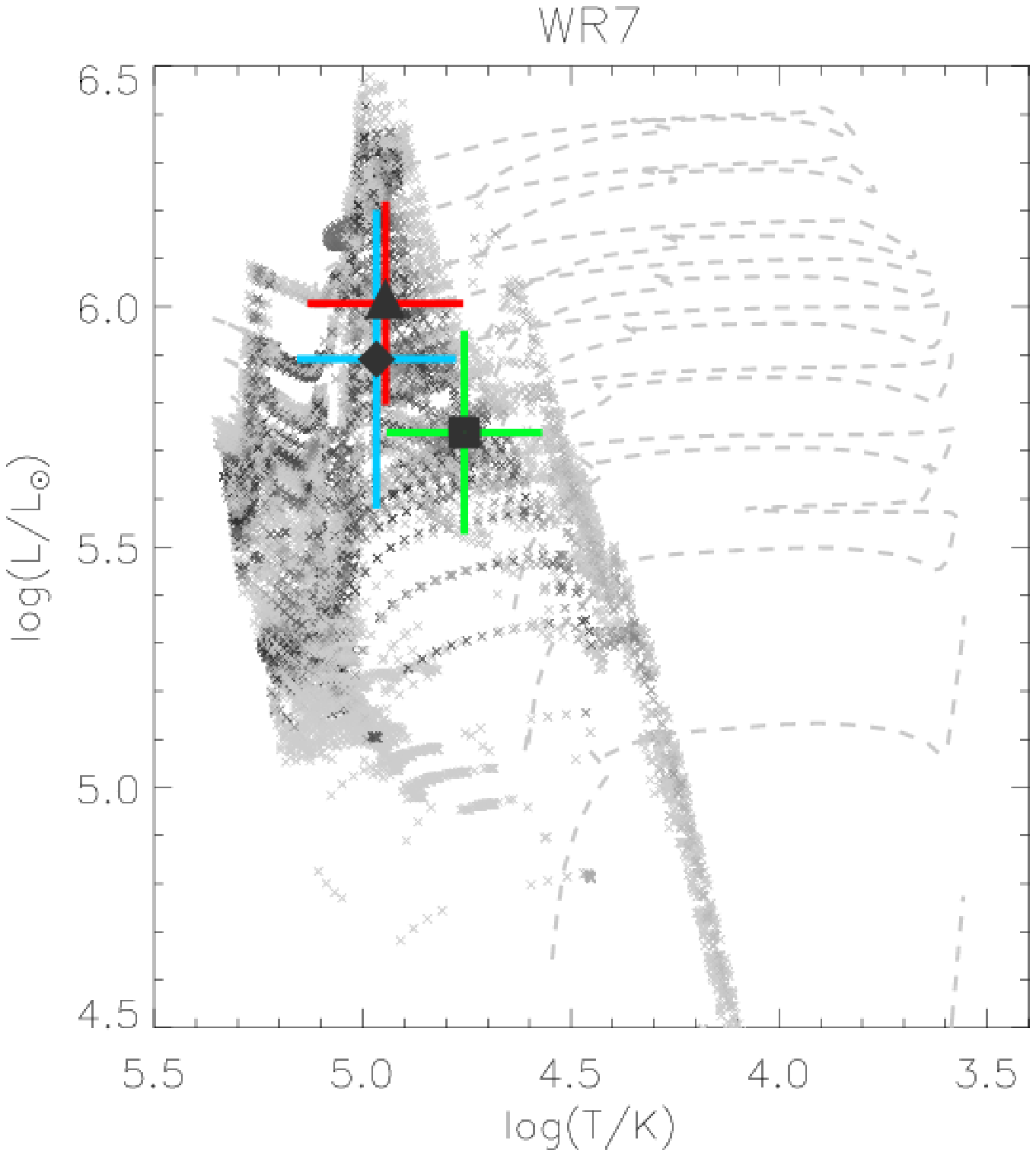}
\caption{Examples of the region on the theoretical Hertzsprung-Russell
  diagram where models match the observed SEDs of RSG1 and WR7. We
  include two plots for RSG1 showing the two fits with and without
  timestep weighing, RSG1a and RSG1b respectively. The dashed grey
  lines indicate single star stellar models with masses of 10, 20, 30,
  40, 50, 60, 70, 80, 100 and 120$M_{\odot}$. The grey points indicate
  the regions where a SED match is made, the darker the point the
  better the agreement between the observed and model SED. The square
  with green error bars indicates the SED fit with the highest value
  of $m \Delta t(t) p_n(m,t)$ from both the single star and binary
  models. The triangle with red error bars indicates the mean $T_{\rm
    eff}$ and $L$ estimated from the single models and the diamond
  with blue error bars indicates the mean $T_{\rm eff}$ and $L$
  estimated from binary models.}
\label{examples}
\end{figure*}

\begin{figure*}
\includegraphics[angle=0, width=58mm]{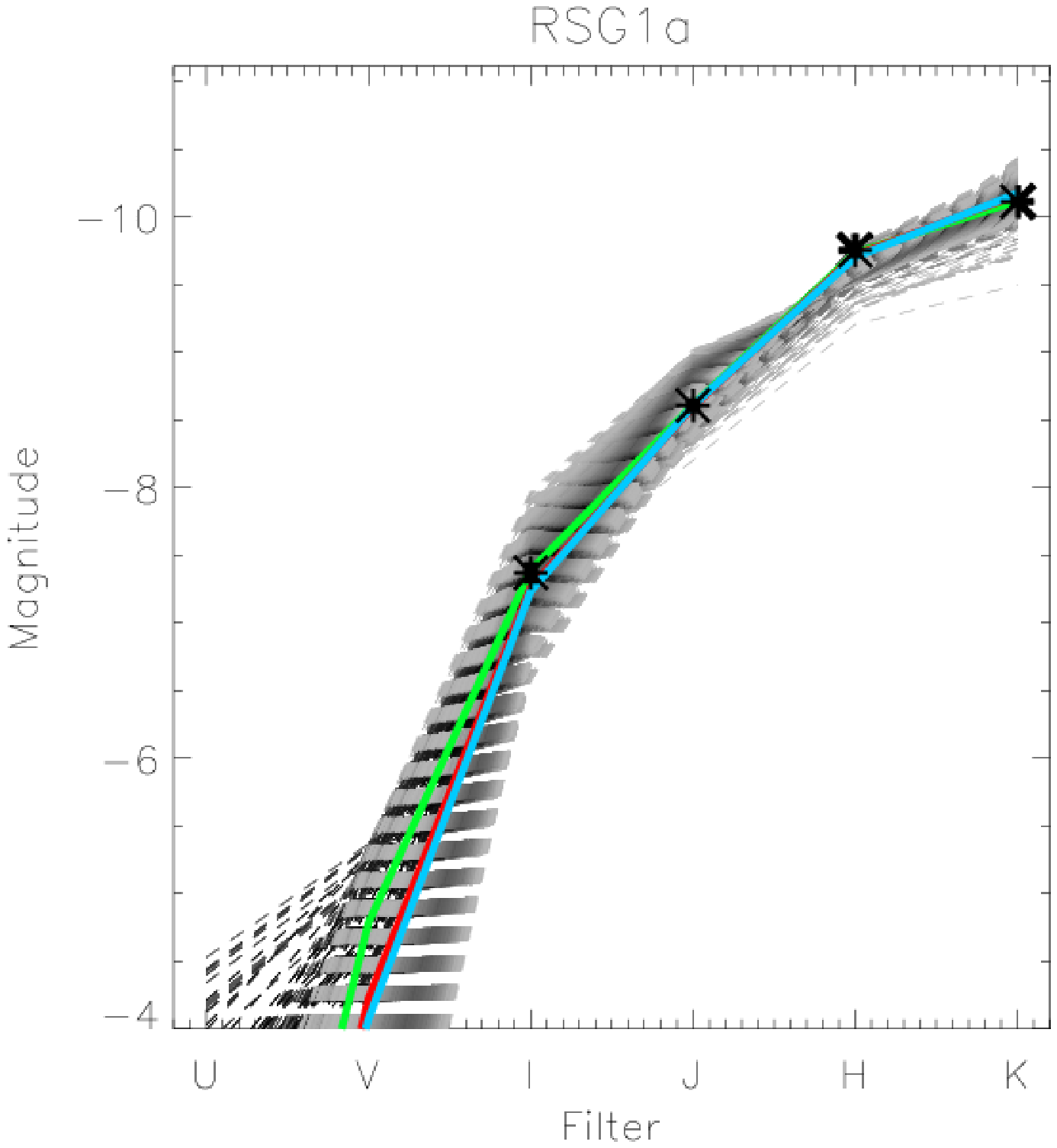}
\includegraphics[angle=0, width=58mm]{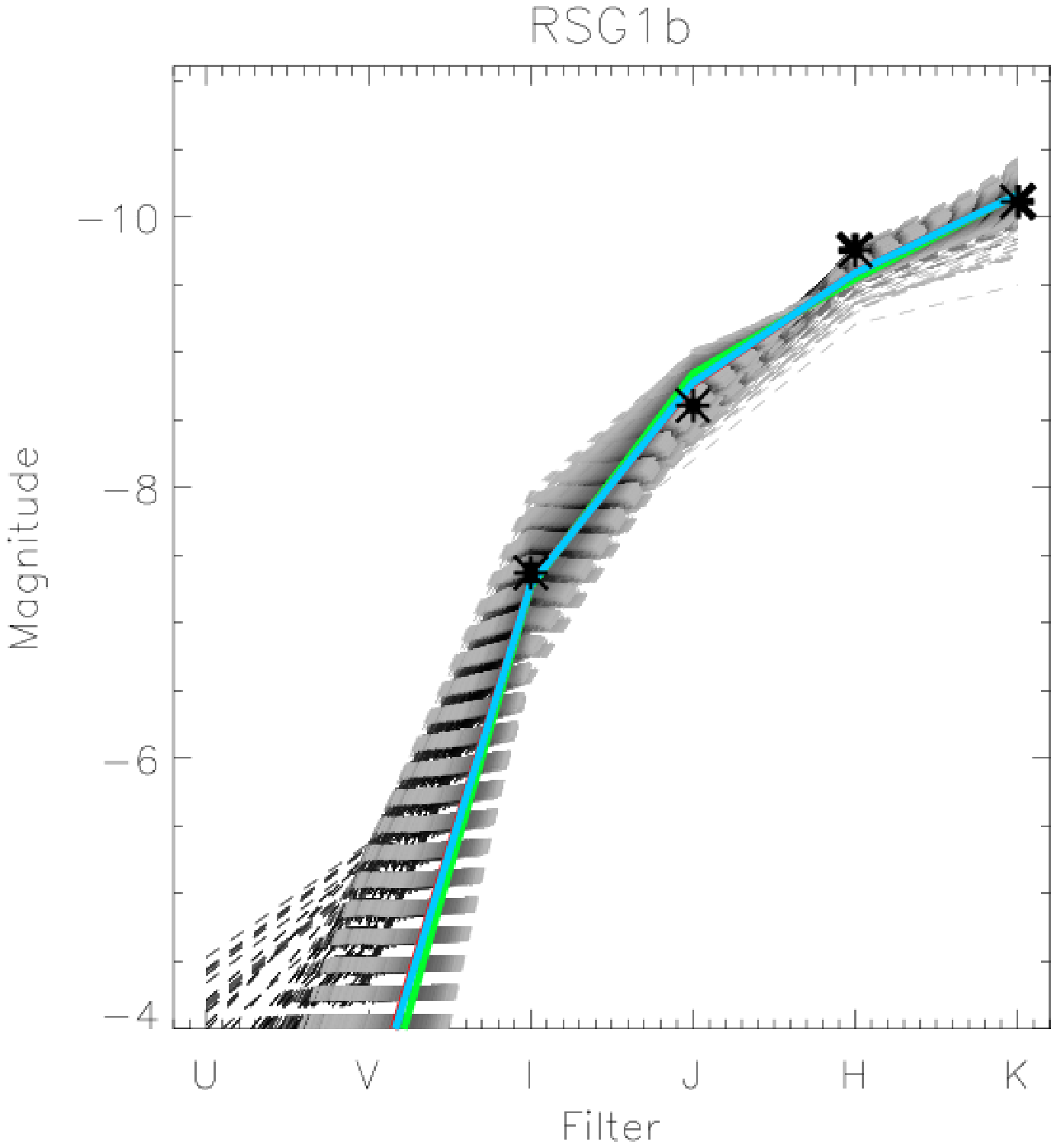}
\includegraphics[angle=0, width=58mm]{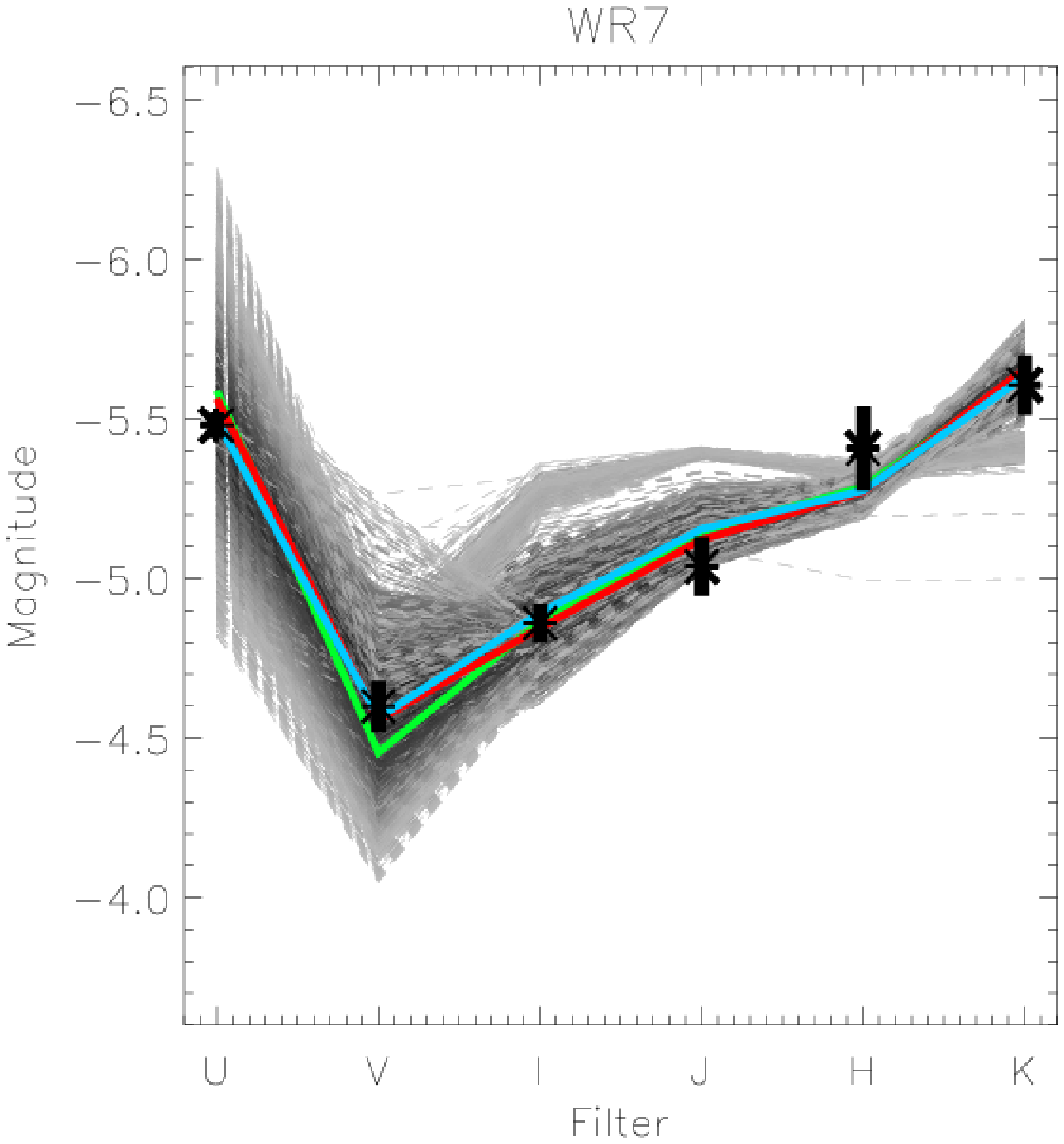}
\caption{Comparison of the observed and theoretical SEDs for the two
  solutions of RSG1 and WR7. The dashed grey lines are theoretical
  matches to the observed SED, the darker the line the higher the
  probability of a match. These lines are the SEDs for the grey points
  in Figure \ref{examples}. The green line is the theoretical SED with
  the highest value of $m \Delta t(t) p_n(m,t)$. The red line
  indicates the mean SED estimated from the single models and the blue
  line indicates the mean SED estimated from binary models. These
  three lines correspond to the colours points in Figure
  \ref{examples}. The asterisks represent the observed SED, the
  vertical lines represent the uncertainty in the observed
  magnitudes.}
\label{sed}
\end{figure*}

In Figure \ref{examples} we show examples of the location of possible
SED matches on the Hertzsprung-Russel diagram.  Then Figure \ref{sed}
compares the theoretical SEDs from the same fits to the observed SED.
For the red supergiant the surface temperature is tightly
constrained but the luminosity is less well constrained. The more
luminous models require more extinction to achieve the same SED and so
tend towards higher surface temperatures. However, despite this large
apparent degeneracy, a low luminosity around $10^5 L_{\odot}$ is
preferred because of the consideration of the timestep. While there
are many possible values of the luminosity, there is only one place
where the star may be observed for an appreciable period of
time. Therefore the fitting process gives greatest weight to this
region and therefore gives the masses and ages listed in Table
\ref{obsstars} below. In this Figure we also include a similar plot
with the timestep weight removed from the SED fit. Without the
timestep weight a better match is achieved by the more luminous and
hence more massive RSG.

We note that in Figure \ref{examples} we predict more massive RSGs
than other stellar evolution models such as \citet{mm03}. However the
lifetimes of these massive RSGs are very short due to their high
mass-loss rate and so they are unlikely to be observed. Furthermore
the evolution of these massive RSGs is also uncertain and extra
pulsation driven mass-loss may shorten their lifetimes further
\citep{yooncanti}.

For the WR star the region on the surface temperature is less well
constrained. This is not surprising as these stars are hot and with
the filters used in the SED fitting we are only looking mainly at the
Rayleigh-Jeans tail of the black-body spectrum so there is little
constraint on the surface temperature. Therefore the SED fits of the
WR and main-sequence stars are intrinsically less well
constrained. Furthermore in Figure \ref{sed} the agreement between the
observed and theoretical SEDs is slightly poorer, especially in the J
and H filters. This is due to most of the WR flux being output at
shorter wavelengths so the IR flux may be more greatly effected by IR
emission from other nearby sources.

An important check to understand the accuracy of the SED fitting
process is to feed theoretical models through the SED fitting code and
compare the output to known inputs. The models we use to do this where
calculated separately to have initial masses between the mass grid of
our stellar models. We use single star models with initial masses of
65, 90 and 100$M_{\odot}$. We also include a $150M_{\odot}$ model to
understand how stars above our maximum model mass would appear. We
calculate fits to the models at three evolutionary phases, the
main-sequence, RSG phase and the WR phase. The results of these fits
are shown in Table \ref{interstars}.

The fits on the RSG phase are most accurate as it is possible to
achieve a better match for cooler stars. This is because the peak of
black-body spectrum is observed rather than for hotter stars where
only the Rayleigh-Jean tail of the black-body spectrum is
covered. Thus both the main-sequence and WR fits are less
accurate. However it is possible to explain the inaccuracy of the the
65 and 80$M_{\odot}$ fits. The the $65M_{\odot}$ SED was taken from
towards the end of evolution for the star while the $90M_{\odot}$ SED
was taken from the near beginning of the track. Therefore there is
some degeneracy in the fitting of main-sequence stars between mass and
age. In addition the more massive stars, including the $150M_{\odot}$
do tend to have their masses underestimated due to this mass-age
degeneracy.  Only with more information from the UV of the stars could
we achieve a more accurate fit. An important conclusion here is that
while individual fits will be uncertain by using the properties of all
of the stars some of these errors will cancel out to provide a
reasonable estimate of the mass and age of NGC~604.

This test indicates that the information inferred on the RSGs can be
trusted more than that on the hotter stars. Also that while each
individual fit may not give the exact result, if we take the entire
population of stars into consideration our combined uncertainty in the
total age and mass of a region is significantly reduced.

\begin{table}
 \caption{Results of test of SED fits for stellar models not in our
   standard grid. The evolutionary phases are, MS main sequence, RSG
   - red supergiant and WR - Wolf-Rayet.}
  \begin{tabular}{@{}cccc}
    \hline
    \hline
Model & Single  &    Binary &\\ 
mass &   star fit &  fit   &   Evolutionary\\
  $/M_{\odot}$ &  $M_i/M_{\odot}$ &  $M_i/M_{\odot}$& Phase\\
\hline
65  & $85\pm 26$ & $ 79\pm  25$ &MS \\
65  & $60\pm 5 $ & $ 55\pm   9$ &RSG \\
65  & $97\pm 21$ & $ 83\pm  29$ &WR \\
90  & $63\pm 15$ & $ 54\pm  17$ &MS \\
90  & $98\pm 18$ & $103\pm  19$ &RSG \\
90  & $90\pm 20$ & $ 85\pm  28$ &WR \\
110 & $89\pm 25$ & $ 98\pm  24$ &MS \\
110 & $88\pm 14$ & $ 89\pm  21$ &RSG \\
110 & $85\pm 25$ & $ 80\pm  31$ &WR \\
150 & $87\pm 24$ & $ 85\pm  25$ &MS \\
150 &$106\pm 10$ & $104\pm  17$ &RSG \\
150 & $85\pm 25$ & $ 80\pm  31$ &WR \\
\hline
\hline
\end{tabular}
  \label{interstars}
\end{table}
 
\begin{figure*}
\includegraphics[angle=0, width=164mm]{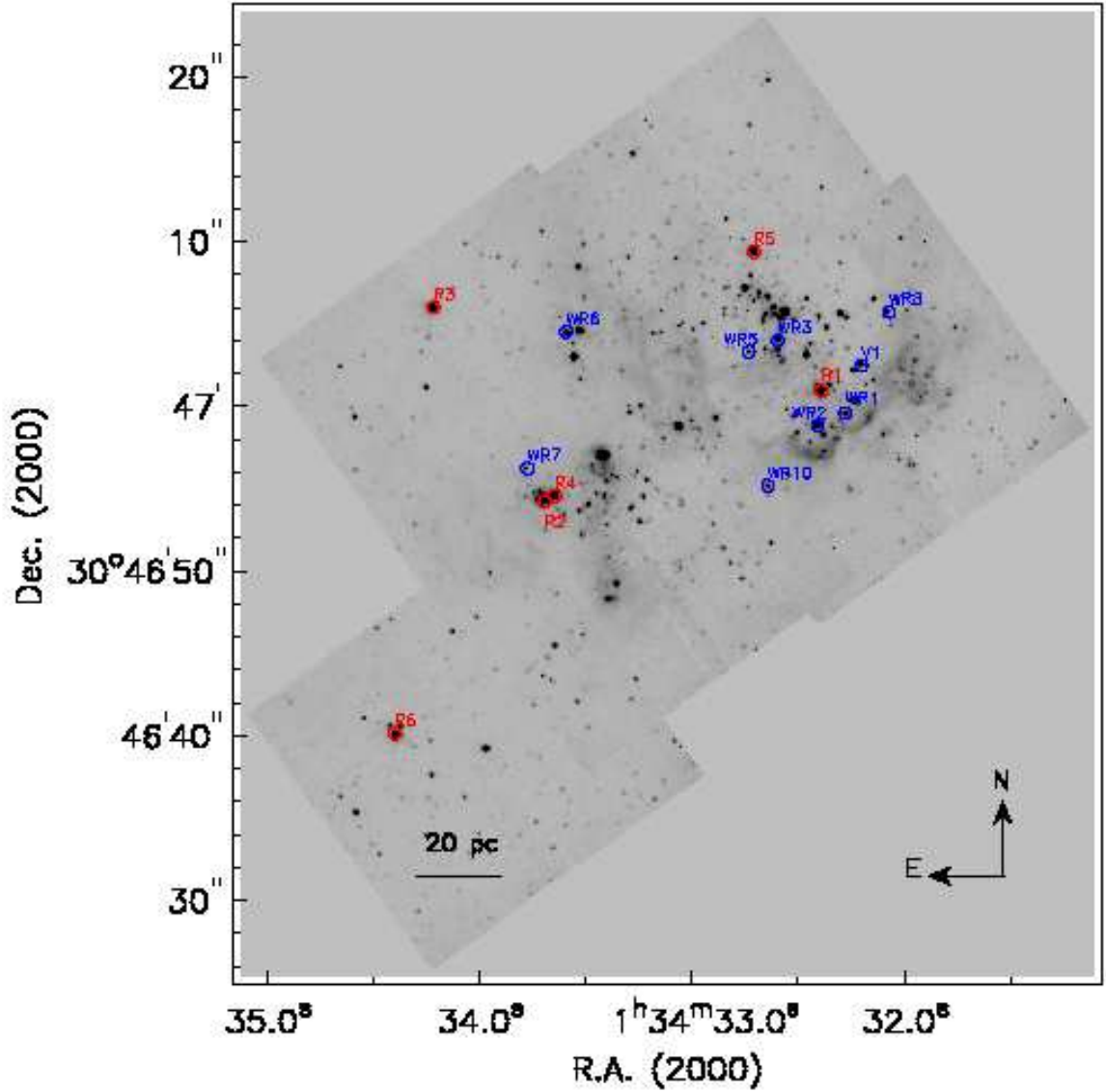}
\caption{Final mosaic F110W image of NGC~604 with the WR stellar
  population (blue circles) from \citet{hunter} and the RSG stars (red
  circles) identified in NICMOS data.}
\label{F205W}
\end{figure*}

Another problem with our fit is due to weighing by the timestep a star
in a rapid phase of evolution may be misidentified as a lower mass,
more slowly evolving, star. Therefore we removed the $\Delta t$ from
the fitting process. We found that the resulting values were similar
to those when the timestep is included. The only exception was
RSG1. In Table\,\ref{obsstars} we show that the mass of the RSG
changes from $23\pm20\rm\,M_{\odot}$ to $67\pm36\rm\,M_{\odot}$ with a
corresponding drop in age and increase in extinction. Such a massive
RSG has a very short evolutionary timescale which is why it is not
favoured when the timestep weight is included. It is not possible from
SED fitting alone to determine which mass is the best solution even
though the more massive, shorter lifetime, RSG fit is less
probable. To determine its true identity a spectrum of the star is
required. However we assume its has a similar age to the other RSGs in
our analysis of the region and we note that its location
(Fig.\,\ref{F205W}) is close to the centre of NGC~604 and suggests
that the more massive solution is possible.

\begin{table*}
 \caption{Results of the SED fitting of the most massive RSG and WR
   stars in NGC~604. We also show the results for the most luminous
   stars in the Hunter's catalogue. Note that RSG2 and RSG5 may also
   be foreground stars and we list the two possible solutions to the
   SED fit for RSG1.}
  \begin{tabular}{@{}lcccccccccccc}
    \hline
    \hline
  &  & & &  & & & $q$ & Spectral \\
 ID &  $M/M_{\odot}$ & Age/Myr & $A_V$ & $\Delta A_v$ & $\log(L/L_{\odot})$ & $\log(T/K)$&  $=M_2/M_1$& Type \\
\hline
 RSG$1_a$ &    $23\pm20$  & $11.5\pm 3.2$& $2.7\pm1.9$ & 2.5 & $5.1 \pm0.4$ & $3.61 \pm0.09 $ & - \\
 RSG$1_b$ &    $67\pm36$  & $ 5.7\pm 5.1$& $6.4\pm2.4$ & 6.2 & $5.9 \pm0.5$ & $3.8  \pm0.2  $ & $0.5\pm0.3$ \\
  RSG3 &       $14\pm2$   & $15.0\pm 2.0$& $1.1\pm0.4$ & 0.3 & $4.8 \pm0.1$ & $3.59 \pm0.02 $ & $0.8\pm0.1$ \\
  RSG4 &       $20\pm3$   & $10.2\pm 1.0$& $3.8\pm0.4$ & 3.0 & $5.2 \pm0.1$ & $3.58 \pm0.02 $ & - \\
  RSG6  &      $16\pm5$   & $12.9\pm 1.8$& $2.7\pm0.6$ & 1.5 & $5.0 \pm0.1$ & $3.59 \pm0.04 $ & - \\
\hline
  S2? & $101\pm24$  & $ 3.8\pm 4.5$& $3.6\pm0.8$ & 2.7 & $6.3 \pm0.4$ & $4.3 \pm0.1 $ & $0.6\pm0.3$ \\
  S5? &  $93\pm19$  & $ 3.2\pm 0.7$& $3.5\pm0.2$ & 3.3 & $6.3 \pm0.2$ & $4.2 \pm0.1 $ & $0.7\pm0.3$ \\
\hline
WR1  &  $100 \pm21 $ &$  3.3 \pm   0.7 $&$  0.8 \pm  0.4$ & 0.5 &$  6.0\pm   0.2 $&$   5.0 \pm    0.2 $& $0.8\pm0.2$ &WCE \\
WR2A &  $ 95 \pm25 $&$   3.4 \pm   0.7 $&$  1.1 \pm  0.3 $& 0.8 &$  5.9 \pm  0.3  $&$  4.8  \pm   0.3 $& $0.8\pm0.2$ &WN \\
WR2B &  $ 78\pm 26   $&$ 3.6  \pm  0.7 $&$  0.2 \pm  0.2 $&-0.1 &$  6.0 \pm  0.2 $&$   4.6  \pm   0.3 $& $0.7\pm0.3$ &WN \\
WR3  &  $ 94\pm 26  $&$  3.3  \pm  0.6 $&$  1.1\pm   0.3 $& 0.9 &$  6.0\pm   0.2  $&$  4.8  \pm   0.3 $& $0.8\pm0.2$ &WN \\
WR5  &  $ 38 \pm20  $&$  6.3 \pm   2.0 $&$  0.4\pm   0.4 $& 0.2 &$  5.4\pm   0.3  $&$  4.8  \pm   0.3 $& $0.4\pm0.3$ &WC6 \\
WR6  &  $ 80 \pm20  $&$  3.4 \pm   0.7 $&$  1.6\pm   0.3 $& 1.3 &$  6.1\pm   0.2  $&$  4.4  \pm   0.1 $& $0.5\pm0.3$ &WNL \\
WR7  &  $ 79\pm 32   $&$ 3.9 \pm   1.2 $&$  1.3 \pm  0.4 $& 0.3 &$  5.9\pm   0.3  $&$  5.0  \pm   0.2 $& $0.4\pm0.3$ &WC4 \\
WR8  &  $ 57 \pm32   $&$ 4.9\pm    1.6 $&$  0.2 \pm  0.3 $&-0.2 &$  5.7 \pm  0.3  $&$  4.7  \pm   0.3 $& $0.5\pm0.3$ &WN \\
WR10 &  $ 67\pm 34   $&$ 4.5\pm    1.5 $&$  0.2 \pm  0.3 $&-0.2 &$  5.8 \pm  0.3  $&$  4.8  \pm   0.3 $& $0.5\pm0.3$ &WN6 \\
VI&     $ 92\pm 26  $&$  3.4 \pm   0.7 $&$  0.5 \pm  0.3 $& 0.2 &$  6.0 \pm  0.2  $&$  4.8  \pm   0.3 $& $0.8\pm0.2$ &Of/WNL\\
\hline 
     H38  &  $ 91 \pm26$  & $2.6\pm1.1$  &  $ 0.5\pm0.3$  & 0.3 &  $  6.1\pm0.2 $ &  $  4.6\pm0.2$ &$0.6\pm0.3$\\
     H40  &  $ 80 \pm28$  & $2.1\pm1.6$  &  $ 0.2\pm0.2$  & 0.0 &  $  6.0\pm0.2 $ &  $  4.6\pm0.2$ &$0.5\pm0.3$\\
     H367 &  $102 \pm21$  & $2.8\pm0.6$  &  $ 1.0\pm0.2$  & 0.8 &  $  6.3\pm0.2 $ &  $  4.4\pm0.2$ &$0.6\pm0.3$\\
     H368 &  $ 98 \pm19$  & $2.8\pm0.6$  &  $ 1.0\pm0.3$  & 0.7 &  $  6.3\pm0.2 $ &  $  4.4\pm0.2$ &$0.6\pm0.3$\\
     H369 &  $ 89 \pm26$  & $2.8\pm1.0$  &  $ 0.5\pm0.3$  & 0.3 &  $  6.1\pm0.2 $ &  $  4.6\pm0.2$ &$0.5\pm0.3$\\
     H371 &  $ 91 \pm28$  & $2.2\pm1.2$  &  $ 0.3\pm0.3$  & 0.1 &  $  6.1\pm0.2 $ &  $  4.6\pm0.2$ &$0.5\pm0.3$\\
     H375 &  $ 81 \pm27$  & $2.0\pm1.5$  &  $ 0.2\pm0.2$  & 0.0 &  $  6.0\pm0.2 $ &  $  4.6\pm0.2$ &$0.5\pm0.3$\\
     H376 &  $ 83 \pm28$  & $2.1\pm1.6$  &  $ 0.4\pm0.3$  & 0.1 &  $  6.0\pm0.2 $ &  $  4.6\pm0.2$ &$0.5\pm0.3$\\
     H377 &  $ 85 \pm29$  & $2.6\pm1.9$  &  $ 0.5\pm0.4$  & 0.3 &  $  6.0\pm0.3 $ &  $  4.7\pm0.3$ &$0.5\pm0.3$\\
\hline
\hline
\end{tabular}
 \label{obsstars}
\end{table*}

\section{Red supergiants in NGC~604}

\begin{figure}
\includegraphics[angle=0, width=84mm]{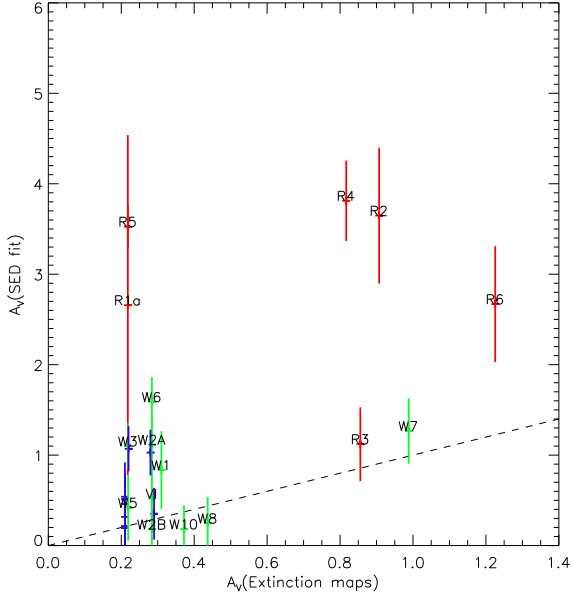}
\caption{Extinction comparison between the values obtained from the Balmer extinction map of 
\citet{relano} and those derived her from the SED fitting. The
  observations are colour codes so that the RSGs are red, WR stars are
  green and main-sequence stars are blue. The names of the RSGs and WR stars
  are given.}
\label{extinction}
\end{figure}

From Fig.\,\ref{mag} we find that there are at least four RSGs within
NGC~604. Determining their ages is vital to understanding the
star-formation history of this region. The results of the SED fitting
are shown in Table\,\ref{obsstars}. We see that, ignoring the unlikely
younger fit for RSG1, the RSGs all have ages in the range of 10 to
15\,Myrs with masses between 14 to 23$\rm\,M_{\odot}$. These are the
typical ages and masses expected for RSGs \citep{rsglev1}. One
important result is that RSGs 1, 4 and 6 have a large extinction
excess above that one predicted from the extinction map derived in
\citet{relano}. In Fig.\,\ref{extinction}, we plot the expected
extinctions of the stars versus the values derived from SED
fitting. Each of these stars is luminous with a bolometric magnitude
in the F205W filter (similar to K-band) of $\approx-10$. Such
extinction excess is found to be typical for RSGs with similar
luminosities \citep{smoke}. The excess is explained as intrinsic
extinction from dust produced in the RSG atmosphere and stellar wind.
A final detail to note is that RSG3 is significantly better fit by a
binary model than a single star.

The second solution for RSG1 with a younger age and higher mass also
has a greater extinction excess. Again this is not unusual when
compared to the excesses found in Galactic RSGs
\citep{smoke}. However, if this solution is true, then this star could
be the most luminous RSG in the local group, as recent observations
indicate that the maximum observed luminosity for typical RSGs is
around $\log(L/L_{\odot}) \approx 5.3$ \citep{rsgl5.3}. Although due
to its higher surface temperature it would probably be classified as a
yellow supergiant (YSG). Its mass and age would be consistent with the
hotter stellar population described below. However, only a spectrum of
the object will truly confirm its nature, as it is the case for RSG4,
the star spectroscopically confirmed as a RSG in \citet{terlevich}. 

The two impostors RSGs S2 and S5 are most likely to be foreground
stars as discussed above. The SED fit gives masses for the stars
similar to the WR and OB stars in NGC~604. Their large extinctions are
similar to the magnitude of the RSGs so they may be transition objects
either becoming WR stars or leaving the main-sequence to become the
RSGs. Only a spectrum for these objects can reveal their true
nature.

In summary, the mean age of the RSGs 1, 3, 4 and 6 is
$12.4\pm2.1$\,Myrs and the mean initial mass is $18\pm4M_{\odot}$. The
other two suspected RSGs are not RSGs but probably foreground stars.

\begin{figure}
\includegraphics[angle=0, width=84mm]{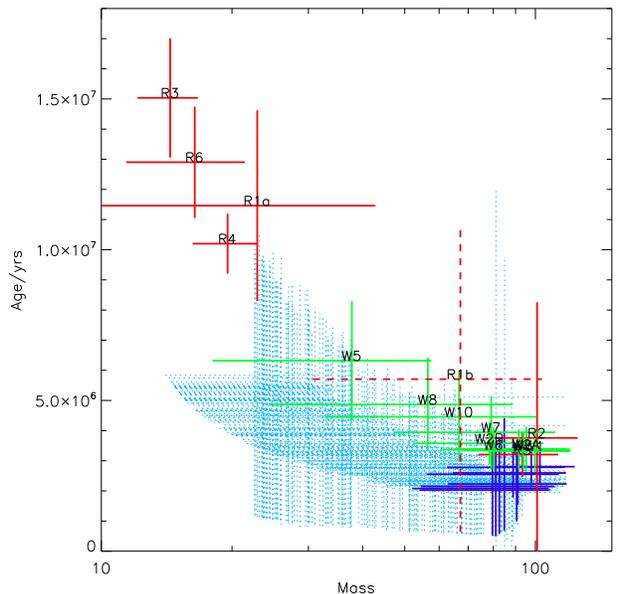}
\caption{The age and mass values derived from the SED fitting for the
  RSGs (red), WR stars (green) and main-sequence stars (blue). The
  lighter blue points are based on UVI SED fitting alone. The names of
  the RSGs and WR stars are given.}
\label{massage}
\end{figure}

\begin{figure}
\includegraphics[angle=0, width=84mm]{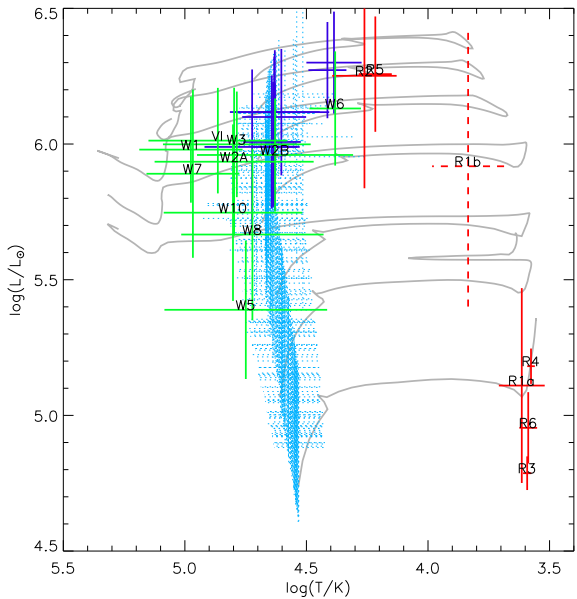}
\caption{The theoretical Hertzsprung-Russell diagram for NGC~604. The
  single-star models in grey have masses of 20, 30, 40, 60, 80, 100,
  120$\rm\,M_{\odot}$. Colour code is: red for RSGs, green for WR
  stars and blue for main-sequence stars. The lighter blue points are
  based on UVI SED fitting alone, names for the RSGs and WR stars are
  given. The R1b star is the second more luminous fit for RSG1. It
  also has a slightly higher surface temperature. Its location on the
  Hertzsprung gap means if this is a solution the stars is in a rapid
  phase of evolution.}
\label{theoryhr}
\end{figure}

\section{Wolf-Rayet stars in NGC~604}

In NGC~604 there are a number of spectroscopically confirmed WR stars
\citep{ngc604wr}. From the available photometry we are able to perform
a SED fit for 10 of these objects and the results are shown in
Table\,\ref{obsstars}. The derived ages and masses are remarkably
consistent, with only WRs 5 and 8 being significantly different from
the mean mass and age of all the WR stars. For each of the stars there
is also a best fit with massive companions with a mass ratio of 0.5 to
0.8.

The WR stars have little excess extinction. We plot the expected
extinction of the stars versus the values obtained from SED fitting in
Fig.\,\ref{extinction}.  We see that the WR stars correspond with the
extinction derived in \citet{relano}.  This is to be expected as any
dust around the star created during its brief RSG phase will be
quickly swept away by the fast and dense WR winds
\citep[e.g.][]{Egrb}. The case with the highest extinction, WR6, is
identified as a WNL star. There stars are thought to have still some
of their remaining hydrogen envelope and therefore will not have
completely removed all the dust from their immediate vicinity. This is
consistent with the evolutionary scenario of a star being a RSG before
becoming a WNL star.

The consistent ages and masses of the fits to the WR stars suggest
that they were all formed in a single and recent burst of
star-formation with a mean age of $4.0\pm1.0$\,Myrs. This is
significantly different to the RSG ages shown above but it is also
similar to the age inferred for RSG2 and RSG5 and the alternative fit
for RSG1. The mean initial mass for the WR stars is $76\pm10M_{\odot}$.

\section{Main-sequence stars in NGC~604}

In the photometric catalogue of \citet{hunter} there are
1040 stars with UVI photometry. We have performed a similar process of
SED fitting for the entire catalogue using these three
filters. Because most of these stars are main-sequence objects the
inclusion of JHK photometry does not increase the accuracy of our
fit. However, for a sample of the most luminous stars in the catalogue
we have also performed a 6 filter fit and listed the results in
Table\,\ref{obsstars}.  The estimated parameters agree within the fit
errors. From the fit of the entire catalogue, we find 34 O stars with
masses above $70\rm\,M_{\odot}$, with the most massive star
$\approx100\rm\,M_{\odot}$. For these 34 stars we find a typical age
of $2.4\pm0.3$\,Myrs. Below $70\rm\,M_{\odot}$ the error in the age of
the SED fit increases significantly, as shown in Fig.\,\ref{massage},
because of the degeneracy between mass and age for O stars. For
example above $60\rm\,M_{\odot}$ we find 53 stars with a mean age of
$2.8\pm1.1$Myrs.

In most cases of the OB stars the derived extinction is similar to
that seen across the NGC~604 region as derived by \citet{relano}. This
is as expected with the OB star winds being fast and sweeping out the
region around each star.

In both mass range mentioned above the ages are slightly younger than
the ages derived for the WR stars but still overlapping within the
errors. This is to be expected because both the OB and WR stars come
from the same initial mass range but are respectively earlier/later in
their relative lifetimes. We created synthetic cluster models of
NGC~604 formed in a bursts lasting between 1 to 2 Myrs. We find that
in these models the true age of the cluster is between the ages of the
OB and WR star populations with the same initial masses. In addition
the duration of the burst is similar to the spread of the two ages of
the two populations.

It is worth noting that the apparent spread also has another possible
explanation the stars might have different rotation rates with those
rotating more rapidly having longer main-sequence lifetimes. If this
is the case then the duration of the star-formation burst would be
decreased.

\section{Stellar population of NGC~604}

From the analysis shown above we see that NGC~604 is a collection of
main-sequence stars, RSGs, possible BSGs and WR stars. We find two
distinct populations: an older one with an age of $12.4\pm2.1$\,Myrs
represented by the RSGs, and a younger one which is a combination of O
stars, possible massive RSG/BSGs and WR stars with a mean age of
$2.4\pm0.3$\,Myrs for the O stars and $4.0\pm1.0$\,Myrs for the WR
stars. We take the age of this younger population to be
$3.2\pm1.0$\,Myrs by combining these last two measurements assuming that the
true age of the burst is intermediate between the relatively \textit{young} O and
\textit{old} WR stars. Importantly the WR stars indicate the upper age for the
star formation burst to be around 5Myrs. This can be visually checked
in Figure \ref{massage} with the ages clustered closely around 2 to 4
Myrs.

We plot our results for the estimated luminosities and surface
temperatures onto a Hertzsprung-Russell (HR) diagram in
Fig.\,\ref{theoryhr}. We see that many of the WR stars cluster around
the track for the $80\rm\,M_{\odot}$, along with many of the most
massive stars and S2 and S5. While RSG 1,3, 4 and 6 are
closer to the $20\rm\,M_{\odot}$ track. This again demonstrates that
the RSGs in NGC~604 are a separate population from the more massive
stars.

For the young and old populations it is possible to place limits on
the total stellar mass formed in each population. Using the mean mass
and its uncertainty to describe the mass range of the stars in each
population, we can estimate the total mass of the stars assuming a
Salpeter IMF (with $\alpha=-2.35$) from 0.5 to $120\rm\,M_{\odot}$ and
a slope with $\alpha=-1.3$ between 0.1 and $0.5M_{\odot}$
\citep{kroupa02}. This only gives us the mass of primary stars,
therefore we multiply the resultant mass by 1.5 to account for each
star having a binary companion. Doing this we assume a flat
distribution of the mass ratio between the two stars with
$0<M_1/M_2\le1$. For the older population with 4 RSGs assuming they
represent all stars between 14 and 22$\rm\,M_{\odot}$ we find a total
mass of $1700\pm900\rm\,M_{\odot}$. For the younger population with
masses above 70$\rm\,M_{\odot}$, we estimate a mass of
($3.8\pm0.6$)$\times 10^5\rm\,M_{\odot}$. We note that if we use all
stars with masses above 60$\rm\,M_{\odot}$ we obtain an identical mass
estimate.

The stellar population of NGC~604 can also be investigated by using
various spectral features in a integrated spectrum for all stars in
the region. \citet{rosa} obtained UV and optical spectra of the
NGC~604 region and studied in detail the nebular emission of the
region. Using their spectrum we have calculated the equivalent widths
for the stellar population diagnostic lines of CIV at 1500\AA, HeII at
1640\AA\ and the Blue WR bump around 4686\AA\ and compared them to
those predicted by our synthetic spectra for the stellar
populations. The calculation of the equivalent widths and the
construction of the synthetic spectra are described in detail in
\citet{es09}.

\begin{figure*}
\includegraphics[angle=0, width=58mm]{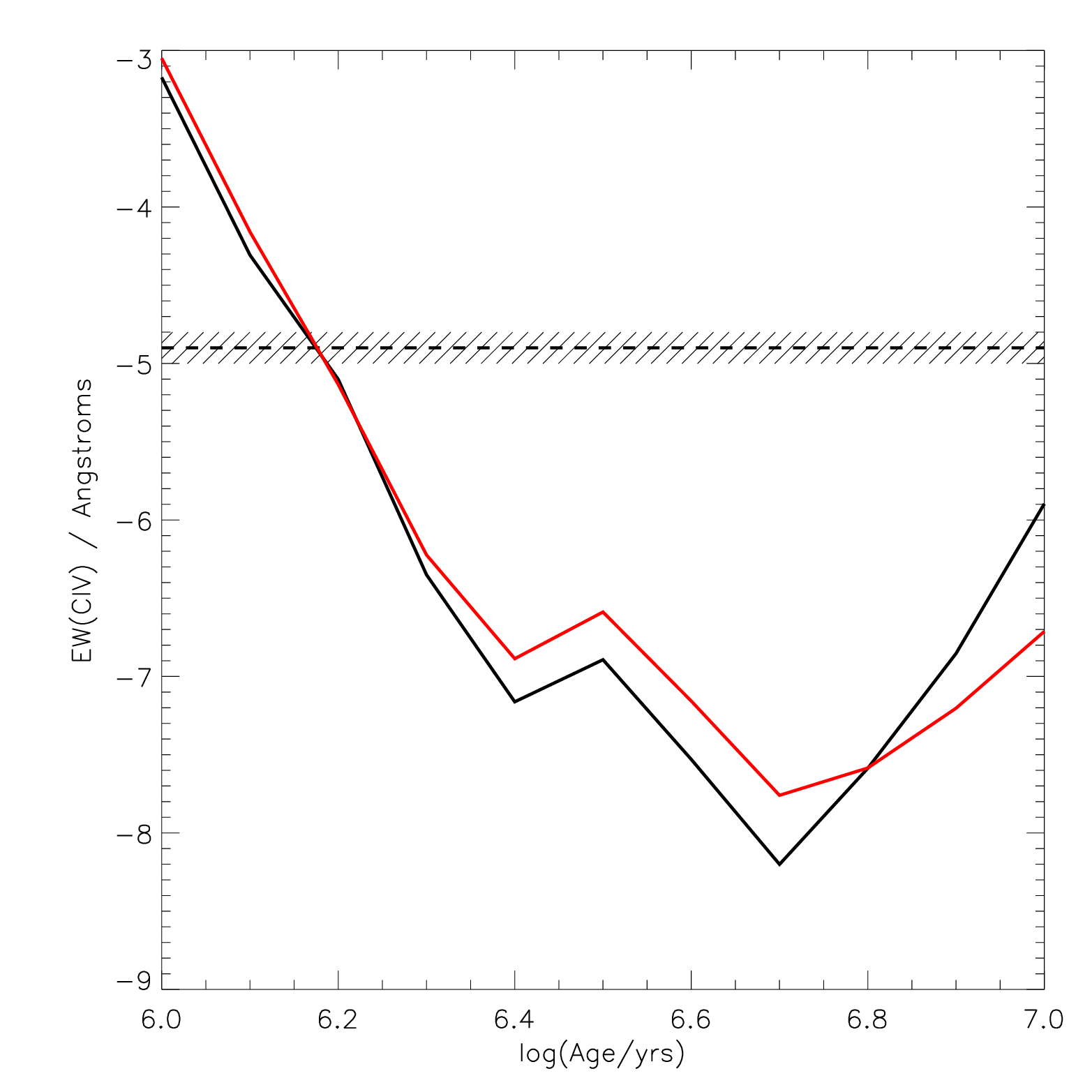}
\includegraphics[angle=0, width=58mm]{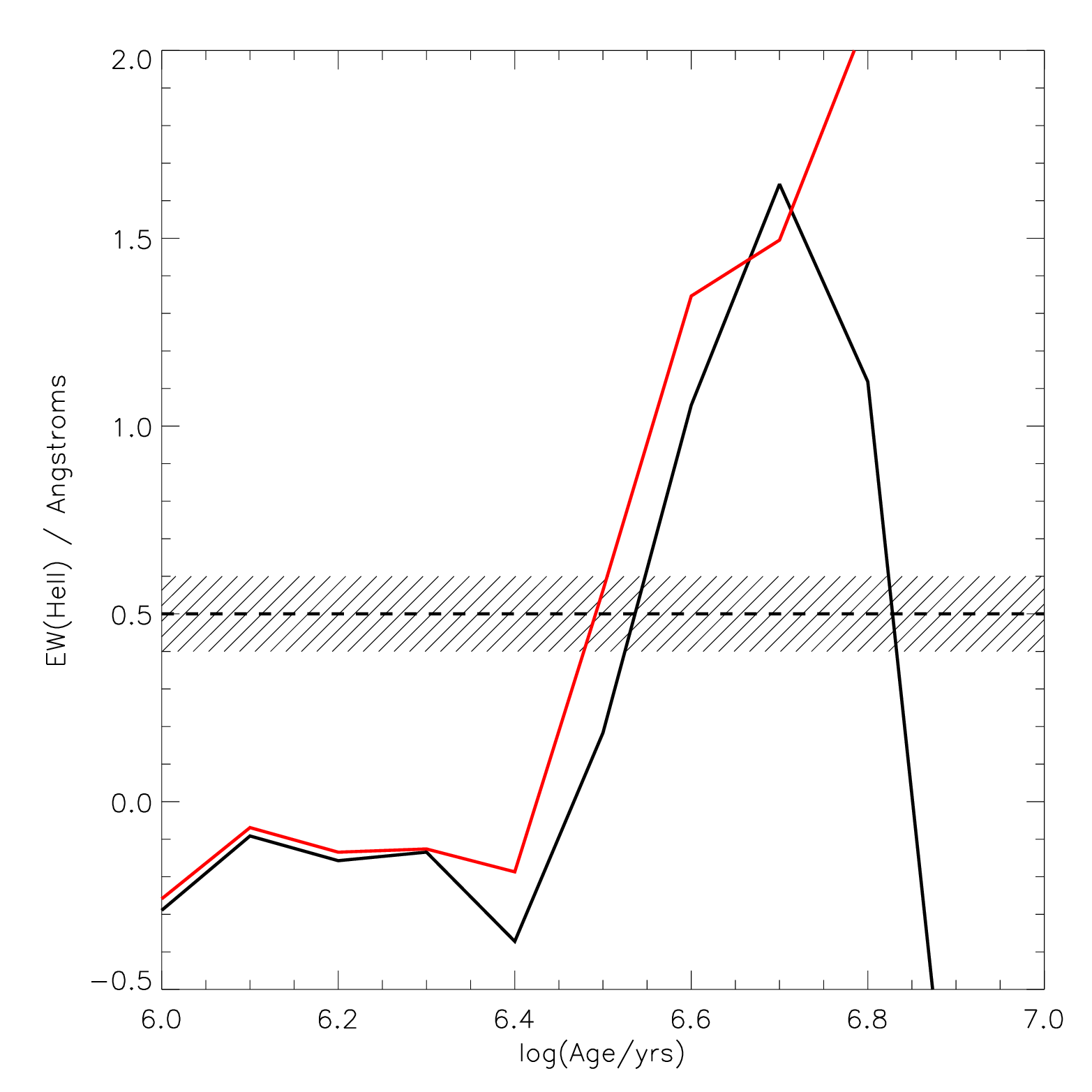}
\includegraphics[angle=0, width=58mm]{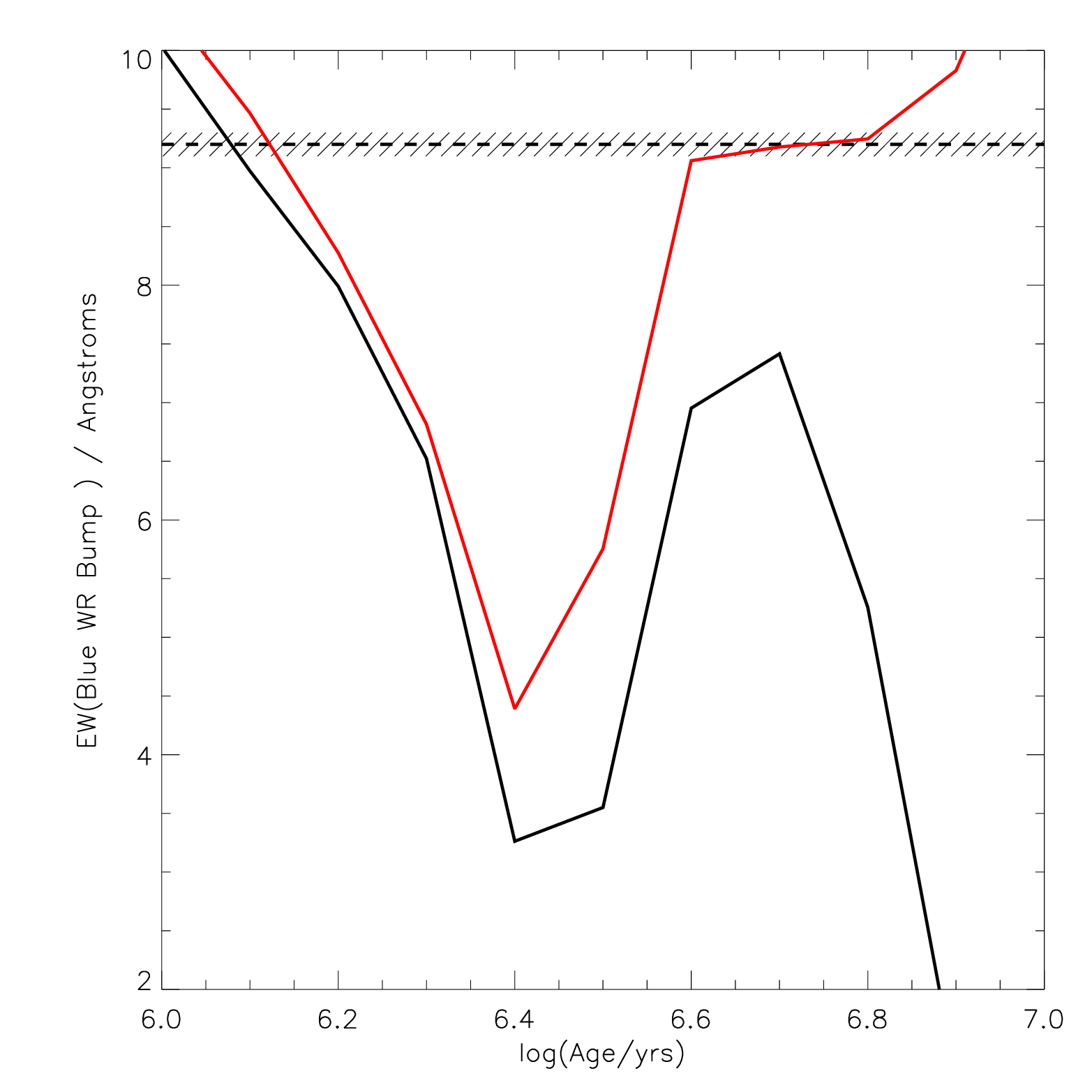}
\caption{The equivalent widths of different stellar population
  diagnostic lines of NGC~604, the dashed lines with errors
  represented by the hashed region, compared to single-star (black)
  and binary (red) population models. Left: the CIV line at
  $1550$\,\AA\ from O-stars, centre: the HeII line at
  $1640$\,\AA\ from WR stars and right: the Blue WR bump around
  $4686$\,\AA.}
\label{ewplots}
\end{figure*}

We plot the comparisons in Fig.\,\ref{ewplots}. We see that the
features that are dependent on the WR stars -the HeII line and Blue WR
bump- are in agreement with the observed values are ages around
3\,Myrs. The CIV line which is dominated by emission from the
main-sequence OB stars has agreement at much earlier ages of
1.5\,Myrs. Therefore we can assume that there is some agreement
between models and the observed spectra but there is no exact
match. We suggest this is because the two spectra do not fully cover
the whole area of NGC~604 and therefore we cannot be sure that the
entire resolved population of stars is contributing to the spectrum
observed. Also the model spectra are for an instantaneous burst where
NGC~604 has an extended period of star-formation. However, the general
agreement provides confidence in the synthetic binary population and
spectral models.

\section{H$\alpha$ and leakage of ionizing photons}

Using the masses for the stellar population given above and the
predicted H$\alpha$ per $\rm\,M_{\odot}$ at different ages listed in
Table\,\ref{halpha} it is possible to estimate the H$\alpha$ flux
these two populations will produce. We calculate the nebula emission
from \textsc{Cloudy} \citep{cloudy} as described in \citet{es09}.  The
ionizing spectrum given to \textsc{Cloudy} as input is obtained using
the fitted stellar populations in this paper (Table\,\ref{obsstars})
and the corresponding stellar interior and atmosphere models. The
predicted flux varies most with the assumed age for the populations.
For the old population we find $\log(F(\rm H\alpha)/{\rm erg \,
  s^{-1}}) \le 36.75$, while for the younger population of
$3.2\pm1.0$\,Myrs we find $\log(F(\rm H\alpha)/{\rm erg \,
  s^{-1}})=39.92^{+0.11}_{-0.20}$. Therefore, it is the younger
population the one which dominates the H$\alpha$ emission from
NGC~604.

The H$\alpha$ flux from NGC~604 has been studied in detail by
\citet{relano}. They obtained an extinction-corrected H$\alpha$ of
$\log(F(H\alpha)/{\rm erg\,s^{-1}})= 39.63\pm 0.07$. Our model
prediction agrees within the errors with this value, but in general
overestimates the H$\alpha$ flux and we investigate this further.

\citet{relano}, using the single-star \textit{Starburst99} code
\citep{starburst} inferred the mass of stars in NGC~604 to be
$5\times10^5\rm\,M_{\odot}$ assuming a Salpeter IMF between 0.1 and
120$M_{\odot}$. This value was derived using single star models and an
age of 4\,Myrs. Using the observed flux, our single star models and an
age of 4\,Myrs we also find a similar mass of
$4.7\times10^5\rm\,M_{\odot}$.  However, the uncertainty in the mass
is greater for single star models because for them the H$\alpha$ flux
varies more with age than for binary star models (see
Table\,\ref{halpha}).  The weaker variation of H$\alpha$ flux for
binaries compared to that from single star models is due to
rejuvenation of secondary stars and mergers from binary evolution.

The binary population models produce more H$\alpha$ flux per stellar
mass over a longer period of time. As most of the stars in
Table\,\ref{obsstars} are better fit with the binary population models
we favour the H$\alpha$ flux prediction from these models. The
difference in the predicted and observed H$\alpha$ flux gives us an
estimate of the leakage fraction of ionizing photons. For NGC~604 we
derive a leakage fraction of $49^{+16}_{-19}$ per cent of the total
ionizing flux in the region. 

This value of leakage is dependent on uncertainties within the stellar
models. Alternative explanations for the mismatch of observed and
predicted H$\alpha$ flux are therefore that either we are
overestimating the number of massive stars or the age of the region
must be greater than 5Myrs. Both these possibilities reduces the
H$\alpha$ flux from the stellar population. However, the large
number of young main-sequence stars and massive WR stars indicates
such an old age is unlikely. Therefore we can be confident in our
results.

\begin{table}
 \caption{Predicted H$\alpha$ flux per $\rm\,M_{\odot}$ of stars for both single star and binary populations.}
  \begin{tabular}{ccc}
    \hline
    \hline
                   & $log(F(H\alpha_{\rm Single})$ & $log(F(H\alpha_{\rm Binaries})$       \\

$\log($Age/yrs$)$  & $/{\rm  erg \, s^{-1}M^{-1}_{\odot}})$ & $/{\rm  erg \, s^{-1}M^{-1}_{\odot}})$    \\
\hline
6.0     &34.34 &  34.41\\
6.1     &34.35 &  34.43\\
6.2     &34.35 &  34.44\\
6.3     &34.35 &  34.43\\
6.4     &34.25 &  34.39\\
6.5     &34.16 &  34.34\\
6.6     &33.96 &  34.20\\
6.7     &33.67 &  34.02\\
6.8     &33.27 &  33.83\\
6.9     &32.75 &  33.54\\
7.0     &32.27 &  33.34\\
\hline
\hline
\end{tabular}
  \label{halpha}
\end{table}

\section{Discussion \& Conclusions}

We have found further evidence that there is a population of RSGs
within NGC~604. This older population inferred from the RSGs
$12.4\pm2.1$\,Myrs. The age of the more numerous young stars is
$3.2\pm1.0$\,Myrs, which is in agreement with previous studies
\citep{maiz,rosa}.

There is some uncertainty in the nature of RSG1 as it may be a very
massive RSG or YSG in a short-lived phase of evolution. To accurately
determine its mass a spectrum of this object is required. However its
deep position in NGC~604 surrounded by many O and WR stars indicates
that it may well be a very massive RSG.

The RSGs have larger extinctions than the other stars in NGC~604. This
is in agreement with them producing large amounts of dust to obtain
greater intrinsic extinction \citep{smoke}. The one WNL WR star also
has a higher extinction than expected from the study of
\citet{relano}. This is in agreement with the evolutionary scenario
that RSGs lose their hydrogen envelopes and become WR stars. The two
stars labeled in Table\,\ref{obsstars} as S2? and S5? may be RSGs in
the very late phases of losing their hydrogen envelopes due to their
very high inferred extinction. However they are more likely to be
foreground stars.

We also find that the WR spectral features HeII and the Blue Bump and
the O-star CIV P-Cygni profile of the composite spectrum of the region
agree in general with the results from the resolved stellar
population. The OB main-sequence stars appear to be younger than the
WR stars. However, a direct comparison is difficult as the UV and
optical spectra of the region do not cover the entire region and the
duration of the star-formation burst is similar to the age of the
cluster and therefore difficult to model.

By using the estimated age of the stellar population we are able to
predict the H$\alpha$ flux from the massive stars and compare this to
the flux observed for NGC~604. We find the two agree in magnitude but
that $49^{+16}_{-19}$ percent of the ionizing flux may
be escaping from the region. The greatest uncertainty in calculating
the H$\alpha$ flux, and therefore the amount of leakage, is the age of
the stellar population. Thus, we are unable to completely rule out
that there are no absolute leakage of ionizing photons. However, since
leakage has been studied to be a common property of luminous HII
regions \citep[e.g.][]{oey97,zurita02,relano02}, it is more likely
that leakage is occurring. Future study of the main-sequence
population using HST/ACS data will enable tighter constraints to be
placed upon the predicted ionizing flux escaping from NGC~604 (Laporte
et al., in prep).

The evidence supporting the existence of RSGs in NGC~604 and their
relation to the most recently formed stars is important because recent
and future cutting-edge telescopes operate at these longer wavelengths
(e.g. Spitzer, Herschel, JWST and the E-ELT). Soon all premier
observatories will operate predominantly at IR wavelengths and
therefore modelling and observation efforts must act now to address
this observational shift. There are known problems with modelling
near-IR emission \citep{maraston,lancon}: typically it is found that
for some star clusters ages derived from near-IR observations do not
agree with those estimated from optical ones.  The RSGs within NGC~604
are a key element required to study distance HII regions in the
IR. The relative contributions of older RSGs and younger stars must be
evaluated. Key to breaking any degeneracy will be the use of spectra
and spectral lines such as the IR lines of RSGs and Wolf-Rayet stars
\citep[e.g.][]{crow2,bend}. This implies that the study of distant HII
regions at IR wavelengths will be challenging unless the contribution
of such contaminants can be fully understood.

\section{Acknowledgements}

We would like to thank the referee Ben Davies for his constructive
comments that have led to a much improved paper. We would like to
thank Rosa Gonz\'alez-Delgado for kindly provides us with the UV and
optical spectra of NGC~604 shown here.  Some of the data presented in
this paper were obtained from the Multimission Archive at the Space
Telescope Science Institute (MAST). STScI is operated by the
Association of Universities for Research in Astronomy, Inc., under
NASA contract NAS5-26555. Support for MAST for non-HST data is
provided by the NASA Office of Space Science via grant NAG5-7584 and
by other grants and contracts. Multidrizzle is a product of the Space
Telescope Science Institute, which is operated by AURA for NASA. This
research was supported by a Marie Curie Intra European Fellowship
within the 7$^{\rm th}$ European Community Framework Programme.

\bsp

\end{document}